\documentclass[aps,floatfix,superscriptaddress,twocolumn,showpacs]{revtex4-1}
\usepackage{amsmath,amssymb,eucal,graphicx}

\def\L{{\rm L}}
\def\R{{\rm R}}

\usepackage{xcolor}
\definecolor{carlos}{HTML}{00A9DE}
\definecolor{lamberto}{HTML}{A900DE}
\definecolor{antonio}{HTML}{FF7F00}

\begin{document}

\title{Heat flux in one-dimensional systems}
\author{Carlos~Mej\'\i{a}-Monasterio}
\email{carlos.mejia@upm.es}
\affiliation{
Laboratory of Physical Properties, Technical University of Madrid,
Av. Complutense s/n 28040 Madrid, Spain}
\author{Antonio Politi}
\affiliation{Institute of Pure and Applied Mathematics, 
Department of Physics (SUPA), Old Aberdeen, Aberdeen AB24 3UE, UK}
\author{Lamberto Rondoni}
\altaffiliation{ORCID:0000-0002-4223-6279}
\affiliation{Department of Mathematical Sciences, Politecnico di Torino,
Corso Duca degli Abruzzi 24, 10129 Turin, Italy}
\affiliation{INFN, Sezione di Torino, Via P. Giuria 1, 10125 Torino, 
Italy}

\begin{abstract}
  Understanding heat  transport in  one-dimensional systems  remains a
  major challenge  in theoretical  physics, both  from the  quantum as
  well as from the classical point  of view. In fact, steady states of
  one-dimensional  systems are  commonly characterized  by macroscopic
  inhomogeneities, and  by long range  correlations, as well  as large
  fluctuations that are typically absent in standard three-dimensional
  thermodynamic  systems.  These  effects violate  locality --material
  properties in the  bulk may be strongly affected  by the boundaries,
  leading  to   anomalous  energy  transport--  and   they  make  more
  problematic the interpretation  of mechanical microscopic quantities
  in terms of thermodynamic observables.  Here, we revisit the problem
  of  heat conduction  in chains  of classical  nonlinear oscillators,
  following  a  Lagrangian and  an  Eulerian  approach.  The  Eulerian
  definition of the flux is composed  of a convective and a conductive
  component.   The   former  component  tends  to   prevail  at  large
  temperatures  where the  system behavior  is increasingly  gas-like.
  Finally, we find that the  convective component tends to be negative
  in the presence of a negative pressure.

\end{abstract}


\maketitle

\section{Introduction}
\label{sec:intro}
A temperature gradient applied to a macroscopic object produces a heat
flow, which  in standard  conditions is  proportional to  the gradient
itself. This is  the content of the phenomenological law  known as the
Fourier  law  of  heat  conduction \cite{Fourier}.  A  great  deal  of
research has been devoted to the  microscopic origin of such a law. In
particular, low-dimensional systems, such as 1-dimensional (1D) chains
of classical  oscillators, have  been targeted  both because  of their
simplicity  and,   more  recently,   also  because   they  approximate
mesoscopic  objects that  are  actually within  reach  of present  day
technology \cite{Lepri:2016}. Despite such  efforts, the derivation of
Fourier law  from microscopic dynamics  remains one of the  major open
problems  of  theoretical  physics \cite{Bonetto}.  Recently,  various
works have  suggested that heat  conduction in  1D systems need  to be
more                        closely                       investigated
\cite{Lepri:2016,LepriSandriPoliti,Hurtado,GRV1,GRV2}, in  view of the
many anomalies that characterise  such systems.  In fact, establishing
meaningful   relationships   between   microscopic   and   macroscopic
properties, primarily  requires accurate  definitions of  the relevant
observables.  In  particular, heat flux  is the crucial  quantity when
the  validity  of Fourier  law  is  to  be investigated.   Irving  and
Kirkwood  provided a  general definition~\cite{Kirkwood},  reported in
many books on  nonequilibrium thermodynamics (see, e.g.,~\cite{Kubo}).
The corresponding expression was derived in Fourier space, where it is
easier  to  establish  its  dependence over  relatively  long  spatial
scales,  those where  hydrodynamic evolution  takes place.   Nowadays,
since numerical  simulations allow accessing  a wide range  of scales,
down  to  the  microscopic  ones,  it is  however,  urgent  to  derive
expressions whose validity is not limited by the spatial resolution.

With this purpose in mind, a real-space version of the Irving-Kirkwood
formula was proposed in Ref.~\cite{LLP}, with reference to 1D systems.
Unfortunately, the definition (23) therein proposed is not as accurate
as initially hoped for. In fact,  in~\cite{GRV1} it is found that in a
stationary regime, the average value of the flux is not constant along
the chain  as it should.   Whether the  sizeable deviations are  to be
attributed to  the strong  chain deformations  observed in~\cite{GRV1}
must be clarified; in any event, Eq.~(23) of Ref.~\cite{LLP}, needs to
be refined.

The task  of this paper is  to revisit the problem,  by distinguishing
different contributions to the heat flux,  with the goal of deriving a
general expression,  valid in 1D  systems both  close to and  far from
thermodynamic equilibrium.  We proceed  by following two different but
equivalent  philosophies:  (i) a  Lagrangian  one,  which consists  in
measuring the  flux as  the energy  exchanged between  two consecutive
particles, irrespective  of where they  are located; (ii)  an Eulerian
one, which consists  in setting a given  (but possibly time-dependent)
threshold and thereby determining the amount of energy that is flowing
through.  The second approach allows for a natural further distinction
between a {\it  convective} component due to  the particles physically
crossing the threshold and a {\it conductive} one, due to the exchange
of (potential) energy  between particles sitting on  opposite sides of
the  threshold  itself.   This   distinction  is  reminiscent  of  the
separation  between the  two  analogous terms  in the  Irving-Kirkwood
formula, the difference being that our quantities concern real space.

The resulting  theoretical formulas are  then tested in two  models: a
Soft  Point   Chain  (SPC),   which  includes  a   confining  harmonic
interaction  and  a short-range  repulsion,  and  a Hard  Point  Chain
(HPC)~\cite{hpc}, characterized by  an infinite-square-well potential.
The  Lagrangian and  Eulerian definition  turn out  to agree  with one
another and  overcome the problem  of Eq.~(23) in  Ref.~\cite{LLP}, in
the sense that the resulting  stationary fluxes are constant along the
chain, as  they should.   Additionally, we explore  the origin  of the
difference between the Lagrangian flux and the conductive component of
the Eulerian flux, as their definitions look formally identical, while
eventually they are not.

It is well known that 1D  systems are neither perfect crystals (in the
thermodynamic limit, particles exhibit arbitrarily large fluctuations,
unless they  are constrained  by an  external substrate),  nor perfect
gases (so long as ordering is  maintained).  We find that the fraction
of   convective-flux  component   is   a  clever   indicator  of   the
``gassiness'' of the underlying system; in  fact, in the limit of very
small  fluctuations  the  conductive component  dominates,  while  the
opposite occurs in  the limit of a ``gas-like" behavior,  such as when
the HPC reduces to a hard-point-gas.

Furthermore and somewhat surprisingly, in both systems, the convective
component of the Eulerian flux tends to be negative in the presence of
a negative pressure,  thus making the conductive part  larger than the
total flux.

Section~\ref{sec:noneq}  is   devoted  to  the  introduction   of  the
formalism    and     to    the    derivation    of     the    relevant
formulae.  Section~\ref{sec:num}  illustrates  the properties  of  the
different flux components  in the two above mentioned  models.  In the
context of  the SPC, we  verify also the relationship  between kinetic
temperature $T$  and density $\rho$ proposed  in Ref.~\cite{GibRon11},
clarifying that  it does not  correspond to  the Boyle law  of perfect
gases.  Moreover,  in Sec.~\ref{sec:hpc}, we prove  a duality property
of  the HPC:  by  denoting  with $a$  the  maximal separation  between
neighbouring  particles, it  turns out  that the  HPC dynamics  for an
average inter  particle distance $\alpha$  is fully equivalent  to the
behavior    of   the    same   model    for   an    average   distance
$\tilde \alpha  = a -\alpha$.~\footnote{Here  and in the  following we
  interchangeably  refer to  $\alpha$  as to  average  distance or  to
  specific   volume.}   Sec.~\ref{sec:pressure}   is   devoted  to   a
discussion of the relationship between the sign of pressure and of the
conductive component  of the heat  flux.  The last section  contains a
summary  of the  main results  and a  brief presentation  of the  open
problems.

\section{Flux definition}
\label{sec:noneq}

Out of equilibrium,  a closed chain of particles in  contact with heat
baths    develops    non-homogeneous   (kinetic)    temperature    and
particle-density profiles, while an energy current flowing from hot to
cold sets in.

In this  section we revisit  the microscopic definition of  the energy
flux, introducing  two different approaches  that in analogy  with the
hydrodynamic description  in fluids,  we define as  ``Lagrangian'' and
``Eulerian''.

For the sake of simplicity  we refer to particle systems characterised
by a  kinetic energy  and nearest-neighbour  interactions, but  we are
confident that the approach herein discussed can be easily extended to
other setups.  In practice, we  assume a one-dimensional system of $N$
interacting  particles,  of  possibly  different  masses  $m_n$,  with
Hamiltonian
\begin{equation} \label{eq:H}
H(\mathbf{q},\mathbf{p},t) = \sum_{n=1}^N
\left [
\frac{p_n^2}{2m_n} + V(q_{n+1}-q_n) 
\right ]  + \zeta_L(T_L) + \zeta_R(T_R) \; ,
\end{equation}
where $\mathbf{q}=(q_1,\ldots,q_N)$  and $\mathbf{p}=(p_1,\ldots,p_N)$
are  the particle  positions  and momentum  vectors respectively,  $V$
denotes the  interaction potential,  and the  $\zeta$ terms  take into
account the  energy exchange between  the system and the  left (right)
heat  bath  at  temperature $T_L$  ($T_R$)  \footnote{The  interaction
  between the  system and  the heat  baths can  be modelled  through a
  deterministic thermostat,  or by  means of  a stochastic  process in
  which case  the dynamics  are Hamiltonian  only in  the bulk  of the
  system.}. Moreover, we explicitly neglect the presence of an on-site
potential,  since it  does not  contribute  to the  flux.  It  should,
however, be reminded that it  indirectly contributes to the scaling of
the  flux  itself  with  the system  size,  determining  whether  heat
transport is normal or not~\cite{LLP}.

The first question  concerns the microscopic definition  of the energy
density, to  represent the  total Hamiltonian as  the sum  of distinct
local contributions $h_n$,  each referring to either  a specific site,
or a  specific link.  Depending  on which  choice is made,  either the
potential, or the kinetic energy  must be (arbitrarily) split into two
different  contributions, attached  to  adjacent sites  or links.   In
spite of such  arbitrariness, no relevant differences  are expected to
emerge  for different  choices over  tens of  microscopic spatial  and
temporal scales. Therefore,  for the sake of  simplicity and symmetry,
we choose to define the local energy $h_n$ as
\begin{equation} \label{eq:Hlocal}
h_n = \frac{p_n^2}{2m_n} + \frac{1}{2}\big [
V(q_{n+1} - q_n) + V(q_{n} - q_{n-1})\big ] \ ,
\end{equation}
assuming  that $h_n$ is localized  on  the position  $q_n(t)$ of  the
particle of interest.

The total Hamiltonian can then be written as
\begin{equation} 
H(\mathbf{q},\mathbf{p},t) = \sum_{n=2}^{N-1} h_n + \Xi_{1;L} + \Xi_{N,R} \  ,
\end{equation} 
where $\Xi_{1;L}$ and $\Xi_{N;R}$ represent  the dynamics of the first
and last particles of the chain and their coupling with the respective
left and right heat baths.

In  1D  chains of  oscillators,  the  energy  flux is  often  computed
referring  to a  specific particle  (or,  better, a  pair of  adjacent
particles), irrespective  of its location,  rather than to  a specific
spatial location.  We start  with this  quantity that,  analogously to
standard hydrodynamics,  we call  Lagrangian, keeping  in mind  that a
flow through consecutive particles only makes sense in 1D systems.

By making use of the equations of motion, the time derivative of $h_n$
can be written as
\begin{equation} \label{eq:continuity}
\frac{dh_n}{dt} = -(j_n^L - j_{n-1}^L) \ ,
\end{equation}
where
\begin{equation}
j_n^L = \frac{1}{2m}(p_{n+1}+p_n)F(q_{n+1}-q_n)  \; .
\label{eq:flux_gen0}
\end{equation}
Eq.~(\ref{eq:continuity}) represents  the energy balance  for particle
$n$,  due to  the energy  flows coming  from the  subsets $(1,n)$  and
$(n+1,N)$  of particles.   It can  also be  interpreted as  a discrete
version of the continuity equation
\begin{equation} \label{eq:continuity-2}
\frac{d\tilde h(\xi,t)}{dt} + \frac{\partial j^L(\xi,t)}{\partial \xi} = 0 \ ,
\end{equation}
where we  have introduced  a pseudo-spatial variable  $\xi =  na$, $a$
being a hypothetical lattice  spacing, or characteristic distance, and
$\tilde h(\xi,t)= h/a$, dimensionally equivalent to an energy density.
At the same time, the flux $j^L$ is dimensionally equal to a 1D energy
density  times a  velocity,  or,  referring to  the  MKS unit  system,
$j^L = \mathrm{J}\mathrm{s}^{-1}$, perfectly  consistent with the flux
estimated from  the interaction with the  heat bath, as the  amount of
energy exchanged per unit time.

This is the only meaningful approach when single oscillators are truly
arranged along a regular lattice  and the variable $q_n$ either refers
to an internal degree of freedom (such as an angle in a spin chain) or
to a  transversal fluctuation.  However, although  almost unnoticed in
the previous literature, it can be  implemented also in the context of
longitudinal fluctuations,  where $a$ does  not need to  coincide with
the   physical   separation   between   consecutive   particles   (see
Sec.~\ref{sec:hpc} where  we show how  the Lagrangian approach  can be
implemented in the HPC).

Let us  now turn  to the  ``Eulerian'' definition  of the  energy flux
through a fixed position $\theta$~\footnote{Here and in the following,
  we assume  $\theta$ to be  constant, but a suitable  time dependence
  can also be  introduced, if deemed useful.}. In this  case, the flux
$j^E_\theta$ is the sum of two contributions,
\begin{equation}
j^E_\theta = j^{D}_\theta + j^{V}_\theta
\label{eq:split}
\end{equation}
where  $j^{D}_\theta$ and  $j^{V}_\theta$, represent  respectively the
conductive and convective component of the flux.  Here, $j^{D}_\theta$
is due to interactions and  it accounts for the (instantaneous) energy
flux from  the $k$-th to  the $k+1$-st  particle, where $k=k(t)$  is a
time-dependent     index,     identified      by     the     condition
$q_k<\theta\le  q_{k+1}$~\footnote{We   implicitly  assume   that  the
  particle  positions are  ordered from  left  to right  and that  the
  collisions do not modify  their order}. The instantaneous expression
of $j^D_\theta$,
\begin{equation}
j^D_\theta = \frac{1}{2m}(p_{k+1}+p_k)F(q_{k+1}-q_k) \;,
\label{eq:flux_1}
\end{equation}
formally    looks     like    the    Lagrangian     flux    expression
Eq.~(\ref{eq:flux_gen0}), with  the difference  that the  particles of
interest change in time for $j^D_\theta$ but not for $j^L_n$.

In turn,  $j^{V}$ accounts  for the physical  motion of  particles: it
represents the  energy flux  due to  particles crossing  the threshold
$\theta$.    This   contribution   to   the   Eulerian   energy   flux
$j^{E}_\theta$, takes  into account  the energy  variation due  to the
particles that  cross the threshold  $\theta$~\footnote{The convective
  component  is  customarily  used  to  compute  the  energy  flux  in
  billiard-like systems of interacting particles and it corresponds to
  the total Euler flux after  requiring that local thermal equilibrium
  conditions  holds \cite{Mejia,Larralde}}.   Since  this  flux has  a
granular structure in  time, it is convenient to refer  it to a finite
time interval $\Delta$  (after all, a flux, as  a macroscopic concept,
takes  a  finite time  to  be  measured).   Therefore, we  define  the
convective component of $j^{E}_\theta$ as
\begin{equation}
j^V_\theta = \frac{1}{\Delta}\sum_{t-\Delta/2<t_j<t+\Delta/2} \! \! \!
\! \! \! h_{k(j)}(t_j) \ \mathrm{sign}[p_{k(j)}(t_j)]
\label{eq:flux_2}
\end{equation}
where the set  $\{t_j\}$ are the discrete times at  which the particle
$k(j)$ crosses  the threshold  $\theta$, and  the sign  function takes
into account the direction of motion.

\section{Numerical analysis}
\label{sec:num}

In this  section, we implement  the above  definitions in a  couple of
relatively  simple  1D  models.   We consider  chains  of  interacting
particles coupled to heat baths  at their boundaries. In particular we
look at  the Hard-Point Chain  (HPC) model, introduced  in \cite{hpc},
that   is  characterised   by  hard-core   attractive  and   repulsive
interactions, and a similar chain model with a soft potential that, in
analogy  to the  HPC, we  call the  Soft-Point Chain  (SPC). We  start
discussing the SPC and then turn our attention to the HPC.

\subsection{Soft-point chain}
\label{sec:spc}

We consider a one-dimensional chain of length $L$, composed of $N=L-1$
particles  with fixed  boundary conditions  (i.e. an  average particle
separation $a=1$) The particles have identical mass $m=1$ and interact
with their  nearest neighbours through  a short-range repulsion  and a
harmonic  attraction.   The Hamiltonian  of  the  system is  given  by
Eq.~(\ref{eq:H}) with potential
\begin{equation}
V(q) = \frac{1}{2}\left( \frac{1}{q^2} + q^2\right) \ . 
\label{eq:mod1}
\end{equation}

\begin{figure}[!t]
\begin{center}
\includegraphics[scale=0.4]{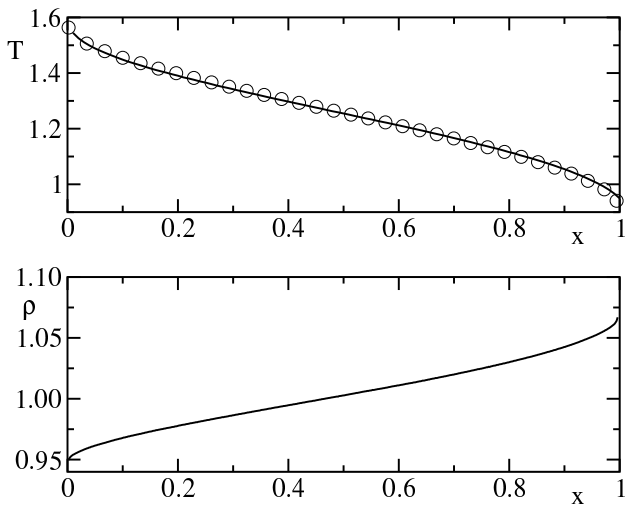}
\caption{Nonequilibrium  densities of  a  soft-point  chain of  length
  $L=512$  coupled   to  heat   baths  at  temperatures   $T_L=2$  and
  $T_R=0.5$. (upper  panel) Temperature profile (solid  curve); (lower
  panel) particle density profile. The circles in the upper panel have
  been               obtained                by               assuming
  $T =5.5 \ (1/\rho- \langle1/\rho\rangle)+\langle T\rangle$.}
 \label{fig:tempdens}
\end{center}
\end{figure}

On  the left  (right) boundary,  an interaction  with a  heat bath  at
temperature  $T_L=2$  ($T_R=0.5$)  is  assumed.   The  interaction  is
simulated by  assuming random collisions with  same-mass particles and
collision  times  uniformly  distributed   within  the  time  interval
$(1,2)$.    The   resulting    temperature-profile   is   plotted   in
Fig.~\ref{fig:tempdens} for a chain of  length $L=512$.  There are, in
principle, two ways  to plot the profile. The first  and most commonly
used consists in adopting the  lattice interpretation, i.e. in setting
$x_n=n/N$ as  the independent variable.  The  second approach consists
in referring to the true  physical position $y$, averaging the kinetic
energy  of the  particle closest  to $y$,  irrespective of  its label.
Here we have adopted the former  approach, but the differences are not
crucial for the messages we want to convey to the reader.

The  abrupt temperature  changes visible  in the  vicinity of  the two
thermal  baths  reveal a  strong  contact  resistance for  the  chosen
parameter  values.  We  expect these  drops to  progressively diminish
when  larger systems  are considered,  since  the number  of modes  of
interaction between baths and system correspondingly increases.

We have  also computed the  particle density profile,  determining the
time averaged positions of particles, $\overline{q}_n$ and the average
inter                         particle                        distance
$s(n) \equiv  \overline{q}_{n+1}-\overline{q}_n$~\footnote{The overbar
  denotes time average.}.  By further  considering that in our setting
the average of $s(n)$ along the chain  is by definition equal to 1, it
makes sense to define the microscopic density as $\rho = 1/s(n)$: this
is the quantity plotted in the lower panel of Fig.~\ref{fig:tempdens}.
Density  and temperature  profiles  are  qualitatively similar.   This
analogy was already noticed in Ref.~\cite{GibRon11} for another model,
where  it  was  suggested  that  $T   =  C_1  \rho^{-1}  +  C_2$,  or,
equivalently,
\begin{equation}
T(x) = C_1 s(x) + C_2 \; .
\label{eq:GR}
\end{equation}
This intuition is here confirmed: the  circles in  the upper  panel   
of Fig.~\ref{fig:tempdens} indeed  correspond to the curve
\begin{equation}
T=5.5 \ (s  -  \langle s  \rangle)+\langle  T\rangle
\label{eq:GR2}
\end{equation}
where the  angular brackets denote  an average along the  chain.  This
relationship may be  obtained from the state equation  of the physical
system, which can  be written as $F(P,s,T)= 0$, where  $P$ denotes the
pressure  and  the  dependence  on  the  volume  is  replaced  by  the
equivalent dependence  on $s$.  Since  the pressure is  constant along
the chain,  a variation  of the temperature  transforms itself  into a
variation  of  $s$ and  this  variation  is,  in  the limit  of  small
displacements,       linear.        So       one       can       write
$F(P,s,T) = F(P,\langle s\rangle+\delta s,\langle T \rangle+ \delta T)
=      F_s     \delta      s     +F_T      \delta     T      =     0$,
which is  nothing but  the formula  proposed in  \cite{GibRon11}.  The
numerically determined coefficient $C_1=5.5$  corresponds to the ratio
$F_s/F_T$, the two derivatives being determined in the middle point of
the profile.  We have thus  identified the constants of that relation,
which had not be done before, and we have confirmed that this relation
does not correspond to Boyle law  of perfect gases (since $C_2$ is not
a small  constant, cf.~\cite{GRV1}), although it  implies that density
is lower at higher temperatures.  Note that Eq.~(\ref{eq:GR}) has been
verified beyond  the small  displacement limit,  which means  that the
conclusions of the above calculation can be generalized by integration
of the infinitesimal variations.

For  what  concerns  the  fluxes,  we  find  that  the  time  averaged
Lagrangian  flux,  $\overline{j}^L_n$,  is   independent  of  $n$  and
approximately equal to $0.171$, while  the time averaged Eulerian flux
is    independent     of    the    threshold     position    $\theta$:
$\overline{j}^E_\theta =  \overline{j}^D_\theta +\overline{j}^V_\theta
\approx           0.084           +           0.087=\overline{j}_n^L$.
Therefore, the two definitions agree with one another, as they should.
Moreover, we see that the  conductive and convective contributions are
approximately  equal  to one  another  for  this choice  of  heat-bath
temperatures. Below,  in this section, we  investigate the temperature
dependence of the two contributions.

\begin{figure}[!t]
\begin{center}
\includegraphics[scale=0.35]{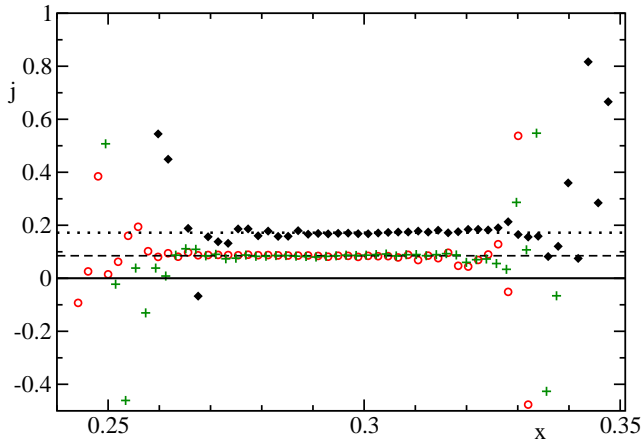}
\caption{(Color  online)  Position  dependence   of  the  energy  flux
  measured with  different procedures in  a chain of length  511 (same
  parameters as in Fig.~\ref{fig:tempdens}).   Black diamonds refer to
  the average Lagrangian  flux $\overline{J}^L_{145}(\xi)$ conditioned
  to  be measured  at the  scaled position  $x=\xi/L$.  Green  plusses
  refer  to  the Lagrangian  flux  of  the  same particle,  under  the
  condition  $x L\in  [q_{145},q_{146}]$.   Red circles  refer to  the
  average conductive  component of the  Eulerian flux measured  in the
  location  $153/L$,  under  the  condition  that  the  label  of  the
  contributing particle is $k$, for  different $k$-values ($k$ is then
  mapped  into  an  $x$-like   variable  via  the  stationary  profile
  $x  = \overline{q}_k/L$).   The horizontal  dashed and  dotted lines
  correspond  to   the  conductive   Eulerian  and   Lagrangian  flux,
  respectively.}
\label{fig:fluxdist}
\end{center}
\end{figure}

Now,  we  discuss  the  relationship  between  $\overline{j}^L_n$  and
$\overline{j}^D_\theta$,  as  they  follow   from  different  ways  of
averaging  the same  quantity  (compare Eq.~(\ref{eq:flux_gen0})  with
Eq.~(\ref{eq:flux_1})).     A    noticeable   difference    is    that
$\overline{j}^L_n$ refers  to a fixed  label $n$, irrespective  of the
position  $q_n$,  while  $\overline{j}^D_\theta$  refers  to  a  fixed
threshold  $\theta$, irrespective  of the  label $k$  of the  adjacent
particles  sitting  across the  position  $\theta$.   It is  therefore
suggestive  to   compute  conditional   averages  to  bring   the  two
definitions closer to one another.  More precisely, we begin computing
$\overline{J}^L_k(\xi)$ as  the (time)  average of $j^L_k$,  under the
condition  that the  center of  mass $Q_k=(q_k+q_{k+1})/2$  is located
inside the  interval $[\xi-d\xi/2,\xi+d\xi/2]$ for a  set of different
fixed positions $\xi$.  The results for  $k=145$ and a chain of length
512  are  shown  in  Fig.~\ref{fig:fluxdist}, (black  diamonds)  as  a
function  of   the  scaled   position  $x=\xi/L$.   We   realize  that
$\overline{J}^L_k$ is independent of  the particle position $\xi$, and
equals the total flux (the fluctuations for relatively small and large
$x$ values  are due to  poor statistics).  One may conclude  that this
comes from  the fact that the  Lagrangian flux is the  total flux, and
that it  does not depend  on the label  of the particle.  However, the
equality  of the  time averages  is not  trivial, as  explained below,
since $\overline{j}^L_n$ and $\overline{J}^L_k(\xi)$ can differ at all
time instants $t$.

Correspondingly,  we  have   computed  $\overline{J}_y^D(k)$,  as  the
average of  $j_y^D$, conditioned  to a set  of preassigned  $k$ values
($k$ denoting the label such that $q_k<\xi<q_{k+1}$).  The results are
shown  in  Fig.~\ref{fig:fluxdist}  (red circles)  for  $x_k=153/512$.
Since $\overline{J}^L_k(y)$ is  a function of the  position $y$, while
$J^D_y(k)$ is a function of the  label, for a meaningful comparison we
have converted $k$ into $y$ by exploiting the knowledge of the average
profile,  i.e.   from  the   knowledge  of  $y=\overline{q}_k$.   From
Fig.~\ref{fig:fluxdist}, we  see that  $J^D_y(k)$ too is  constant and
still equal to the conductive  component of the heat flux.  Therefore,
it is not the combined dependence of the two quantities on $n$ and $y$
to be responsible for the  observed differences between the Lagrangian
flux and the conductive component of the Eulerian flux.

A third kind of conditional average helps to clarify the origin of the
difference.   Given  an index  value  $m$,  as  normally done  in  the
Lagrangian  approach, we  determine the  average of  all instantaneous
flux   values   $j^L_m(t)$,   counting    only   those   events   when
$q_m<y<q_{m+1}$, no  matter how close is  the center of mass  $Q_m$ of
the pair $(q_m$,  $q_{m+1})$ to the assigned threshold $y$  (as it was
done in  the computation  of the diamonds).   We call  this observable
$\tilde    J(m,y)$;   the    results    are    displayed   again    in
Fig.~\ref{fig:fluxdist}  for $m=145$  and  different threshold  values
(see  plusses).  Once  again this  flux  is independent  of where  the
threshold is located.  Less trivial is  that the outcome of this third
type of  protocol now coincides  with the conductive component  of the
Eulerian  flux.   We  can  therefore  conclude  that  the  subtle  but
important  property which  is responsible  for the  difference between
$\overline{J}^L_k(y)$ and  $\tilde J(k,y)$ is  that in the  first case
the average is restricted to those  moments when the center of mass is
close to a given threshold, while in the second case, it is the matter
of the threshold to be contained within the interval $[q_k,q_{k+1}]$.

\begin{figure}[!t]
\begin{center}
\includegraphics[scale=0.35]{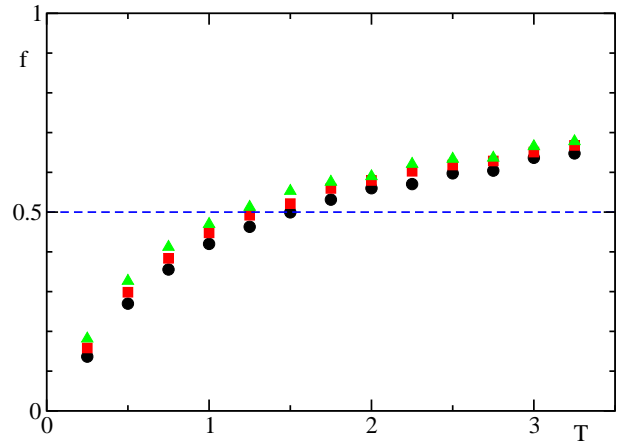}
\caption{(Color                    online)                    Fraction
  $f   =\langle  j^V\rangle/\langle   j^L\rangle$   versus  the   mean
  temperature  $T=(T_L+T_R)/2$,  for  three different  chain  lengths:
  $L=256$  (black dots),  $L=512$ (red  squares), and  $L=1024$ (green
  triangles).  In all cases  $T_L-T_R=0.2$. The dashed horizontal line
  corresponds  to   an  even   distribution  between   conductive  and
  convective component.}
\label{fig:ratio}
\end{center}
\end{figure}
 
Having verified  that the definitions  given in the  previous sections
are meaningful, it  is instructive to look at their  relative size for
different temperatures. In Fig.~\ref{fig:ratio}, we plot the fraction
$f =\overline{j}^V/\overline{j}^L$
versus  the average  temperature $T=(T_L+T_R)/2$  for three  different
chain lengths.   This way, the scaling  behavior of the flux  with the
system  size  does  not  matter  and  we  can  easily  identify  which
contribution prevails.  By definition, $0\le  f\le 1$, the two extrema
corresponding to a  purely conductive ($f=0$) and  a purely convective
($f=1$)  flux.   From  Fig.~\ref{fig:ratio},  we  see  that  at  small
temperatures, the  convective components  is relatively  negligible in
the SPC.  In fact, it is  reasonable to conjecture that $f(T)$ goes to
zero for $T\to  0$, since the fluctuations of the  particle around the
equilibrium  position decrease  and so  does the  number of  threshold
crossings  which contribute  to  the convective  flux.  A  preliminary
analysis suggests  that the convective contribution  vanishes linearly
with $T$.  On  the other hand, at larger  temperatures, the convective
component  becomes  dominant,   reflecting  the  wilder,  increasingly
gas-like behavior  of the  particles along  the chain.   Moreover, the
ratio $f$ slowly  grows with $L$ at fixed $T$.   It probably saturates
for $L\to\infty$, but more detailed numerical studies are necessary to
test this hypothesis.

Finally, we  look at temporal  fluctuations.  In order to  average out
the irrelevant microscopic  fluctuations, it is convenient  to look at
the total  flux.  In the  case of the  Lagrangian approach, it  is the
matter of averaging  $j^L_n$ over all $n$ values.  In  order to have a
statistically equivalent  definition of the Eulerian  contribution, we
have determined  it for a  set of equispaced thresholds,  separated by
the average inter particle distance.

\begin{figure}[!t]
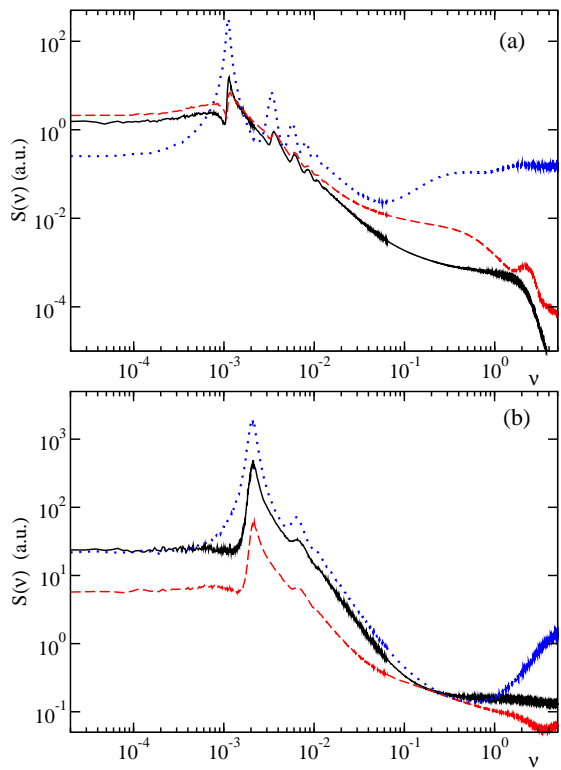

\begin{center}
\includegraphics[scale=0.3,clip=true]{fig.4a.eps}
\includegraphics[scale=0.3,clip=true]{fig.4b.eps}
\caption{(Color online) Power spectrum of  the heat flux for $L=1024$:
  panel  a   (b)  refers   to  $T_L=0.3$  ($T_L=3.3$)   and  $T_R=0.2$
  ($T_R=3.2$).   The different curves correspond to the Lagrangian
  (black solid), Euler  conductive (red dashed),   and   Euler
  convective (blue dotted) fluxes.}
\label{fig:fluxspect}
\end{center}
\end{figure}

The  results are  plotted  Fig.~\ref{fig:fluxspect}  for a  relatively
small  and a  large temperature.   The  peak exhibited  by all  curves
correspond  to time  needed by  a sound  wave to  propagate along  the
chain.   Somehow  surprisingly,   stronger  harmonics  components  are
visible at lower  temperatures, when the dynamics  should in principle
be more  sinusoidal. It is also  interesting to see that  the spectral
weight of the convective component prevails also at small temperatures
(see the  upper panel)  where its  average value  is smaller  than the
conductive one (this can be extrapolated  from the height of the power
spectrum at extremely low frequencies).

\subsection{Hard-point chain}
\label{sec:hpc}

We now  study the energy flux  in the HPC.  The  model, introduced and
studied  in~\cite{hpc}, consists  of  a one-dimensional  chain of  $N$
particles of masses $\{m_i\}$,  positions $\{q_n\}$ and linear momenta
$\{p_n\}$ ordered along a  line.  Nearest neighbour particles interact
through elastic  collisions when $q_{n+1}-q_n =0$  (type-A collisions)
and $q_{n+1}-q_n = a$ (type-B collisions) as given by the square--well
potential in the relative distances defined by
\begin{equation}
 V(q_{n+1}-q_n)= \left \{ \begin{array}{lll}
     0 & , &\quad \mbox{$0\;<q_{n+1}-q_n\;<\;a$}\\
     \infty & ,& \quad \mbox{otherwise}
     \end{array} \right.
\label{u}     
\end{equation}
Type-B collisions can be visualised as if the particles were linked by
an inextensible and  massless string of fixed length  $a$.  Both types
of collisions  are of hard-core  type, and  are described by  the same
rule.  Referring to the pair  $(n,n+1)$, the particles' momenta change
as
\begin{eqnarray} \label{eq:collisions}
p^\prime_n & = & \frac{m_n-m_{n+1}}{m_n+m_{n+1}} p_n +
                 \frac{2m_n}{m_n+m_{n+1}} p_{n+1} \\
p^\prime_{n+1} & = & \frac{2m_{n+1}}{m_n+m_{n+1}} p_n
                     -\frac{m_n-m_{n+1}}{m_n+m_{n+1}} p_{n+1}
\end{eqnarray}
where the primed momenta correspond to their values after a collision.
To avoid ballistic energy transport, here we consider a diatomic chain
for which the masses of  the particles alternate between two different
values that we  chosen as $m_{n}=1$ for even  $n$ and $m_{n}=\sqrt{2}$
for odd $n$.

The chain has  fixed boundary conditions, meaning that for  a chain of
length $L$, we include two  ``virtual'' particles with fixed positions
$q_0=0$  and  $q_N=L$.  The  chain  length  sets the  specific  volume
$\alpha=L/N$, which  is constrained to be  $0<\alpha<a$ and determines
the prevalence of type-A versus type-B collisions.  The chain internal
pressure $P$ is obtained as the  average change of momentum due to the
collisions, namely $P_n = p_n^\prime -  p_n$.  It is easy to note from
Eq.~(\ref{eq:collisions})  that the  pressure  is  independent of  the
particle index,  and thus  homogeneous with  respect to  the position.
The internal  pressure $P$ is  positive for $\alpha<a/2$  and negative
for $\alpha>a/2$  \cite{hpc}.  In  what follows,  and without  loss of
generality, we set the maximal particle distance to $a=1$.
 
The  stationary  nonequilibrium  state  is  set  by  thermalising  the
particles next to the heat baths,  namely $q_1$ and $q_N$, by means of
a  stochastic  process.   Each  thermalized  particle  bears  its  own
exponential  clock;  as the  clock  ticks  the corresponding  particle
acquires a new velocity drawn from an equilibrium thermal state at the
corresponding   temperature   ($T_\L$   or  $T_\R$).    We   set   the
thermalisation   rate   to   $10^3$.   In   between   collisions   and
thermalisation, the particles move according to the HPC rules.

\begin{figure}[!t]
\begin{center}
\includegraphics[scale=0.7]{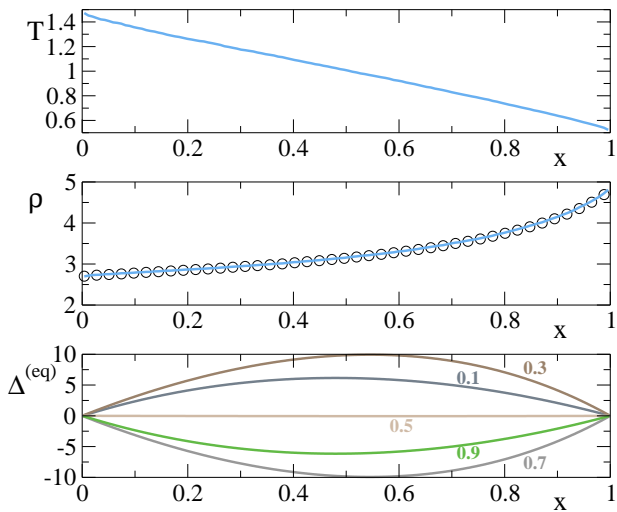}
\caption{(Color online) Nonequilibrium densities of a hard-point chain
  consisting of $N=512$ particles, with temperature difference
  $T_L=1.5$, $T_R=0.5$ and $\alpha=0.3$.  (upper panel) Temperature
  profile $T(x)$; (middle panel) particle density profile $\rho(x)$
  (solid curve).  The circles correspond to the particle density
  obtained from the temperature profile through the equation of state
  Eq.(\ref{eq:hpc-EOS}), with $P=2.4$ numerically computed as the
  momentum change due to collisions.  (lower panel) Deformation of
  the chain $\Delta^{(\mathrm{eq})}$ for different values of $\alpha$ (as
  indicated by the labels).
  \label{fig:hpc-dens}}
\end{center}
\end{figure}

Temperature   and    particle   density   profiles   are    shown   in
Fig.~\ref{fig:hpc-dens}  for  $\alpha=0.3$.  The  temperature  profile
(upper panel) is similar to the  one obtained for the SPC, though here
we do not observe temperature discontinuities at the contacts with the
heat baths.   This is just  a consequence of the  stronger interaction
assumed   herein.   However,   given  that   the  properties   of  the
nonequilibrium  state  are  determined   by  the  bulk  dynamics,  the
existence of  a contact  resistance is irrelevant.   We also  show the
particle  density  (middle panel),  computed  as  the inverse  of  the
average inter  particle distance  $\rho$, which  is equivalent  to the
number of particles per unit length. The equation of state for the HPC
was derived in \cite{hpc}, and is given by
\begin{equation} \label{eq:hpc-EOS}
\rho(x)  =  \frac{T(x)}{P}  -  \frac{1}{\mathrm{e}^{P/T(x)}  -  1} \ ,
\end{equation}
where $P$ is  the internal pressure.  The circles in  the middle panel
of  Fig.~\ref{fig:hpc-dens}, obtained  through Eq.~(\ref{eq:hpc-EOS}),
show  an  excellent  agreement.   It is  worthwhile  noting  that  the
equation of state  governing the HPC local state differs  from that of
the SPC  Eq.~(\ref{eq:GR}), except at  small deviations from  the mean
density and kinetic temperature.

For $T_L \ne T_R$, the  chain deforms inhomogeneously. We measure this
deformation as  the deviation of  the average particle  positions with
respect   to   their  equilibrium   positions   $q_n^{(\mathrm{eq})}$:
$\Delta^{(\mathrm{eq})}(n)  =  \overline{q}_n -  q_n^{(\mathrm{eq})}$.
In   the    lowest   panel   of   Fig.~\ref{fig:hpc-dens}    we   show
$\Delta^{(\mathrm{eq})}$  for   different  values  of   $\alpha$.  The
deformation  vanishes  at  the  border   due  to  the  fixed  boundary
conditions; moreover, it  is either positive or  negative depending on
the sign  of the  internal pressure. For  $\alpha=1/2$, i.e.  for zero
pressure,  $\Delta^{(\mathrm{eq})}=0$ along  the  chain.  Finally,  it
looks like  the maximal deformation  exhibits a peak  for intermediate
specific volumes:  this is  an ''artifact"  of the  interaction scheme
with  the  heat  baths.   In  fact,   for  both  $\alpha  \to  0$  and
$\alpha\to  1$,  the  average   time  separation  between  consecutive
collisions  goes to  zero, i.e.  the time  scale of  the HPC  dynamics
becomes arbitrarily  short.  On the  other hand, the  interaction with
the heat bath being ruled by  a fixed time scale, becomes increasingly
weak.

\begin{figure}[!t]
\begin{center}
\includegraphics[scale=0.35,clip=true]{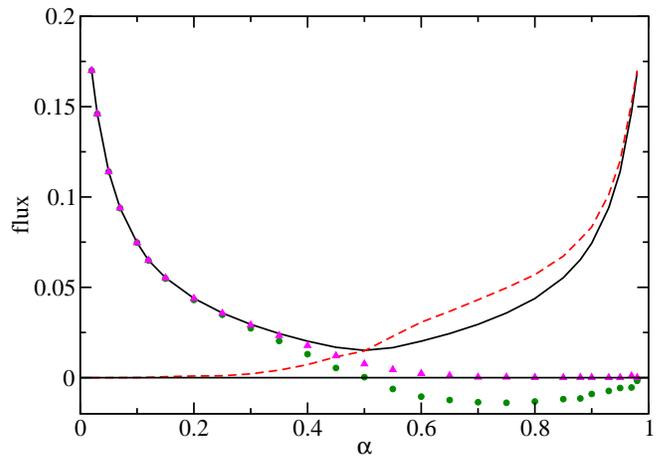}
\caption{(Color online) Different fluxes for the HPC, as functions of the
  specific volume $\alpha$: total flux (black solid curve),
  Euler conductive flux (red dashed curve), Euler convective flux
  (green knots), and the contribution of type-A collisions to the
  flux (magenta triangles). All  simulations are performed for a
  chain of  length 1024, $T_L=1.5$,  and $T_R=0.5$. The  ratio between
  the two  masses is 1.5.  The distribution of interaction  times with
  the heat bath is uniform between 0 and 2 in all cases.}
\label{fig:fluxHPC}
\end{center}
\end{figure}

A similar implication is found with  reference to the energy flux that
we are now  going to discuss.  We have implemented  both the ``Euler''
and     the     ``Lagrange''     description     as     defined     in
section~\ref{sec:noneq}.   Once again,  both  currents  have the  same
value, $j^L=j^E$, and  do neither depend on the particle  index nor on
the physical position, as it must be. The dependence of the conductive
and  convective components  of the  heat flux  on the  specific volume
$\alpha$ is illustrated in Fig.~\ref{fig:fluxHPC}.

There, we observe that the total flux is symmetric as 
a function of the parameter $\alpha$, about its value $1/2$.
This is due to a duality linking the HPC models with specific 
volumes $\alpha$ and $\hat{\alpha}$, when $\hat{\alpha} = 1-\alpha$.
In fact,  given the configuration  I$=\{q_1, q_2, q_3,  \ldots, q_L\}$
with $q_0=0$ and $q_L=\alpha_L$, let us build the sister configuration
II$=\{r_0,r_1, r_2,\ldots ,r_L\}$, starting  from $r_0 = (1-\alpha)L$,
and recursively  defining $r_{m+1} =  r_m - 1  + (q_{L-m+1}-q_{L-m})$.
Since  the original  length  is  $\alpha L$,  the  length  of the  new
configuration is equal  to $(1-\alpha)L$.  Let us  finally assume that
the   velocities  of   the  configuration   II  are   left  unchanged:
$\dot r_{m} = \dot q_{L-m}$.

Consider  now  two  consecutive  particles with  positions  $q_n$  and
$q_{n+1}$  in the  configuration I  and  assume that  they are  moving
against one  another.  They  will undergo a  type-A collision  after a
time $\tau  = d/(\dot q_{n+1}-\dot  q_n)$, where $d =  q_{n+1}-q_n$ is
the initial mutual distance.  Within  the sister configuration II, the
corresponding  particles   of  coordinate  $r_m$,  $r_{m+1}$   sit  by
construction at distance $1-d$ and move  away from one another, as the
ordering has  been exchanged.   Therefore, they  will undergo  a type-B
collision  after the  same  time  $\tau$ as  in  the configuration  I.
Moreover, since the velocities are  the same in the two configurations
they remain  equal after  the collision.   Analogously a  collision of
type-B in the  first configuration is fully equivalent  to a collision
of  type-A in  the second  one.  We  can therefore  conclude that  the
initial relationship  between the two configurations  is maintained at
all future times.  In particular, one must expect that the energy flux
is the same in both setups, as observed in Fig.~\ref{fig:fluxHPC}.

As  a  second  observation,  we  notice that  the  flux  diverges  for
$\alpha \to  0$ (and  then also for  $\alpha \to  1$).  Qualitatively,
this is  because the density  increases and the time  interval between
consecutive collisions correspondingly decreases, tending to 
vanish. One may thus expect the  flux to  diverge as $1/\alpha$.
However, the previously noticed decrease of the 
interaction strength,  hence of the rate of thermalization, slows down 
such divergence.

From  the definition  of the  fluxes in  section \ref{sec:noneq},  one
could naively expect that type-A and  type-B collisions of the HPC are
responsible for the convective  and conductive flux, respectively.  In
fact, for $\alpha \to 0$,  type-B collisions become increasingly rare,
the   dynamics  is   gas-like  and   the  conductive   flux  vanishes.
Analogously,  for $\alpha  \to  1$, type-A  collisions disappear,  the
dynamics  is  crystal  type  and the  convective  component  vanishes.
However,  Fig.~\ref{fig:fluxHPC},  where  the contribution  of  type-A
collisions is reported (see the triangles), reveals that the agreement
with the convective component only holds for $\alpha <1/2$.

For $\alpha >1/2$, the disagreement is not only quantitative, but even
qualitative, since the contribution of type-A collisions stays
positive, while the true convective component becomes even negative.
This somehow surprising behavior is further investigated in the
following section.

\section{Role of pressure}
\label{sec:pressure}

From  Fig.~\ref{fig:fluxHPC},  we  see  that for  $\alpha  >1/2$,  the
convective component  of the  flux is  negative, while  the conductive
component is consistently larger than the total flux to compensate for
the negative  convective contribution.  It is  plausible to conjecture
that in the  HPC the sign of  the convective component of  the flux is
related  to the  sign of  the pressure,  since this  latter observable
changes sign  precisely for  $\alpha=1/2$.  In  order to  test whether
this is the general case, we have run some simulations with the SPC by
varying the specific volume.

\begin{figure}[!t]
\begin{center}
\includegraphics[scale=0.35,clip=true]{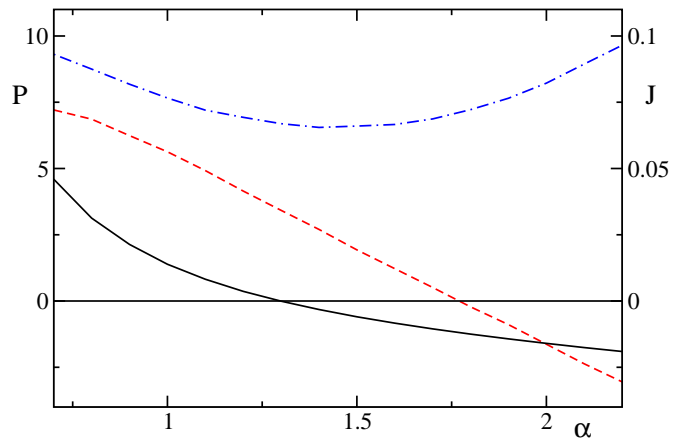}
\caption{(Color  online) Dependence  of  pressure  (black curve,  left
  labels)  in the  SPC  on  the specific  volume  $\alpha$.  The  blue
  dotted-dashed  and red  dashed curves  refer to  the convective  and
  conductive  components   of  the   flux  (see  the   right  labels).
  Simulations have been  performed in a a chain  of $N=255$ particles,
  with  $T_L=1.5$, and  $T_R=0.5$; the  collision times  are uniformly
  distributed in the interval $[0.2,0.4]$.}
\label{fig:fluxSPC}
\end{center}
\end{figure}

In  the  case  of  the   SPC,  when  the  specific  volume  $\alpha=1$
(equivalently  the  average  particle   distance)  the  repulsive  and
attractive  force balance  each  other,  and thus  the  chain at  zero
temperature is at  equilibrium with zero pressure.   Upon switching on
the temperature, we  expect the pressure to increase and  this is what
we see in Fig.~\ref{fig:fluxSPC}, where the solid curve corresponds to
the pressure as a function of the specific volume (see the left axis).
Upon  increasing $\alpha$  we expect  at  some point  the pressure  to
change sign  and this is what  happens for $\alpha \approx  1.29$.  In
parallel to the  computation of the pressure, we  have determined also
$j^V$ and $j^D$,  finding that the convective  component decreases and
eventually changes  sign as in the  HPC.  However, the change  of sign
does not occur at the same point ($j^v =0$ for $\alpha\approx 1.77$).

Therefore,  although  these  simulations  qualitatively  confirm  that
decreasing  the  pressure  contributes   to  decrease  the  convective
component of  the energy  flux, which  eventually becomes  negative, a
quantitative  connection between  the  two observables  is  yet to  be
determined.

\section{Concluding remarks}
In this  paper we have investigated  different microscopic definitions
of  energy flux  in 1D  oscillators systems.   In particular,  we have
considered  two models  (SPC  and HPC)  characterized  by a  repulsive
potential,  which  preserves  the  particles  order  and  a  confining
potential.

In both models, we have found  that the Eulerian and Lagrangian fluxes
are equal  to one another as  it should be. Concerning  the conductive
part $j_\theta^D$,  we have  observed that the  conditional Lagrangian
flux   $\overline{J}^L_k$  equals   the   total   flux,  rather   than
$j_\theta^D$, despite  the formal  similarity of the  two definitions.
The   reason  has   been  found   in   the  fact   that  the   average
$\overline{J}^L_k(y)$ is restricted  to the instants of  time in which
the   center  of   mass  is   close  to   a  given   threshold,  while
$\tilde J(k,y)$ is computed when  the threshold is within the interval
$[q_k,q_{k+1}]$.

At small temperatures, the convective component $j_\theta^V$ is small,
as  one  expects,  but  it becomes  dominant  at  large  tempreatures.
This confirms an increasing similarity with a gas-like behaviour, 
which cannot be perfect if Eq.~(\ref{eq:GR2}) extends to
very high temperatures. We have  also observed a
small increase with  the system size, which is  probably a finite-size
effect, but should be more thoroughly investigated.

The power spectra  of the total flux show a  peak in correspondence of
the inverse of  the time needed by a sound wave  to propagate along the
chain.  Interestingly, stronger  harmonic components  appear at  small
temperatures,  apparently contradicting the idea that the dynamics 
should then resemble  that  of   harmonic  oscillators.  However,  this
peculiarity  might  be  an  instance of  the  long-range  correlations
produced   by   nonequilibrium   boundary   conditions   (see,   e.g.,
Ref.~\cite{GibRon11}).   An additional  peculiarity is  the relatively
large amplitude  of the  spectrum of the  covariant component  even at
small temperatures, when the average of $j^V$ is relatively small.

Rather surprisingly,  in both the SPC  and the HPC, we  found that the
convective component of the Eulerian flux  tends to be negative in the
presence  of a  negative  pressure, thus  making  the conductive  part
larger  than the  total  flux.  At  the moment,  however,  we have  no
indication of a quantitative  relationship between negative conductive
fluxes and negative pressures.

Finally,  we have  have  shown  that the  symmetry  of  the heat  flux
(invariance  under the  transformation $\alpha  \to 1-\alpha$)  in the
HPC, is a consequence of a duality of the model itself.

\vskip 10pt

\section*{Acknowledgements}
LR  has   been  partially   supported  by   Ministero  dell'Istruzione
dell'Universit\`a  e  della  Ricerca (MIUR)  grant  ``Dipartimenti  di
Eccellenza  2018-2022’’.  CMM  thanks the  Department of  Mathematical
Sciences of  Politecnico di  Torino for its  hospitality, acknowledges
financial    support    from     the    Spanish    Government    grant
PGC2018-099944-B-I00  (MCIU/AEI/FEDER,  UE).  This  work  started  and
developed while CMM was a  long term Visiting Professor of Politecnico
di Torino.

\bibliographystyle{apsrev4-1} % Tell bibtex which bibliography style to use
\bibliography{./1D}

\end{document}